\documentclass[useAMS,usenatbib]{mn2e}
\usepackage{myaasmacros}
\usepackage{graphicx}
\usepackage{amssymb}

\DeclareMathAlphabet{\mathitbf}{OML}{cmm}{b}{it}

\title[NGC 5466: a unique probe of the Galactic halo shape ]{NGC 5466: a unique probe of the Galactic halo shape}

\author[H. Lux et al.]{H. Lux$^{1,2}$\thanks{Hanni.Lux@nottingham.ac.uk}, J. I. Read$^{3,4}$, G. Lake$^2$, K. V. Johnston$^5$ \\  $^1$ School of Physics and Astronomy, University of Nottingham, University Park, Nottingham, NG7 2RD, UK\\ $^2$ Department of Theoretical Physics, University of Z\"urich, Winterthurerstr. 190, CH-8057 Z\"urich, Switzerland\\ $^3$ Institute for Astronomy, Department of Physics, ETH Z\"urich, Wolfgang-Pauli-Strasse 16, CH-8093 Z\"urich, Switzerland\\
$^4$Department of Physics and Astronomy, University of Leicester,University Road, Leicester LE1 7RH, UK\\ $^5$ Department of Astronomy, Columbia University, Pupin Physics Laboratories, 550 West 120th Street, New York, New York 10027,\\ USA }
 
\begin{document}
 
\date{Accepted XXX. Received XXX; in original form XXX}

\maketitle

\begin{abstract}
Stellar streams provide unique probes of galactic potentials, with the longer streams normally providing the cleaner measurements. In this paper, we show an example of a short tidal stream that is particularly sensitive to the shape of the Milky Way's dark matter halo: the globular cluster tidal stream NGC 5466. This stream has an interesting deviation from a smooth orbit at its western edge. We show that such a deviation favours an underlying oblate or triaxial halo (irrespective of plausible variations in the Milky Way disc properties and the specific halo parametrisation chosen); spherical or prolate halo shapes can be excluded at a high confidence level. Therefore, more extensive data sets along the NGC 5466 tidal stream promise strong constraints on the Milky Way halo shape.
\end{abstract}

\begin{keywords}
Galaxy: halo, Galaxy: structure, Galaxy: kinematics and dynamics
\end{keywords}

\section{Introduction}\label{sec:intro}

Pure dark matter simulations of galactic halos in our current $\Lambda$CDM paradigm (cold dark matter with a cosmological constant) predict triaxial halo shapes \cite[e.g.][]{1991ApJ...378..496D,2002ApJ...574..538J}. However, if baryonic effects are included, the shape becomes more axisymmetric and aligned with the gas/stellar disc \citep[e.g.][]{1994ApJ...431..617D,2008ApJ...681.1076D,2010ApJ...720L..62K}. Hence, determining the shape of the Milky Way halo constrains both our cosmology, and our current galaxy formation models. The shape of the Galactic potential can also be used to constrain alternative gravity models \citep{2005MNRAS.361..971R}.

Stellar streams are a powerful tool for probing the Milky Way (MW) potential \citep{1977MNRAS.181...59L,2001ApJ...551..294I}. The Sagittarius stream \citep{2001ApJ...547L.133I}, is a textbook example \citep{2004ApJ...610L..97H,2005ApJ...619..800J,2005ApJ...619..807L,2006ApJ...651..167F,2006ApJ...642L.137B,2009ApJ...703L..67L,2010ApJ...714..229L}, but its large width, its bifurcation, and the unknown properties of the progenitor galaxy introduce large systematic errors to the recovered MW halo mass and shape \citep{2010MNRAS.408L..26P,2011ApJ...727L...2P}. Thinner streams avoid these complications (Lux et al. 2012, in prep), but so far have not yielded constraints on the Milky Way halo shape owing to one or more of: close proximity or low inclination with respect to the MW disc; short length; or poor data quality (c.f. \citet{2010ApJ...711...32N} for the Orphan stream and \cite{2009ApJ...697..207W,2010ApJ...712..260K} for the GD1 stream).

Using only angular positions and radial velocities of thin streams, \cite{2011MNRAS.417..198V}  find that the observation of turning points (apo-/pericentres) is crucial for their method to constraint the shape/mass of the probed potential. An example where a potential low apocentre can provide a strong constraint is the globular cluster stream NGC 5466. Independently of each other, two groups reported evidence for this rather tenuous tidal stream \citep{2006ApJ...637L..29B,2006ApJ...639L..17G} that is an order of magnitude fainter than the tidal stream associated with the globular cluster Pal 5 \citep{2003AJ....126.2385O,2006ApJ...641L..37G}. Both groups used the Sloan Digital Sky Survey (SDSS) data but different extraction techniques to identify the stripped cluster stars. \cite{2006ApJ...637L..29B} report a $\sim 4^\circ$ stream using neural networks to extract the probability distribution of the cluster stars, while \cite{2006ApJ...639L..17G} found a $45^\circ$ stream using an optimal contrast matched-filter technique. Because of its limited data set (on-sky positions only)  and the disputed length of the stream the NGC 5466 stream has not been used to constrain the Milky Way halo shape so far. 

\begin{figure*}
\begin{center}
\includegraphics[width=0.92\textwidth]{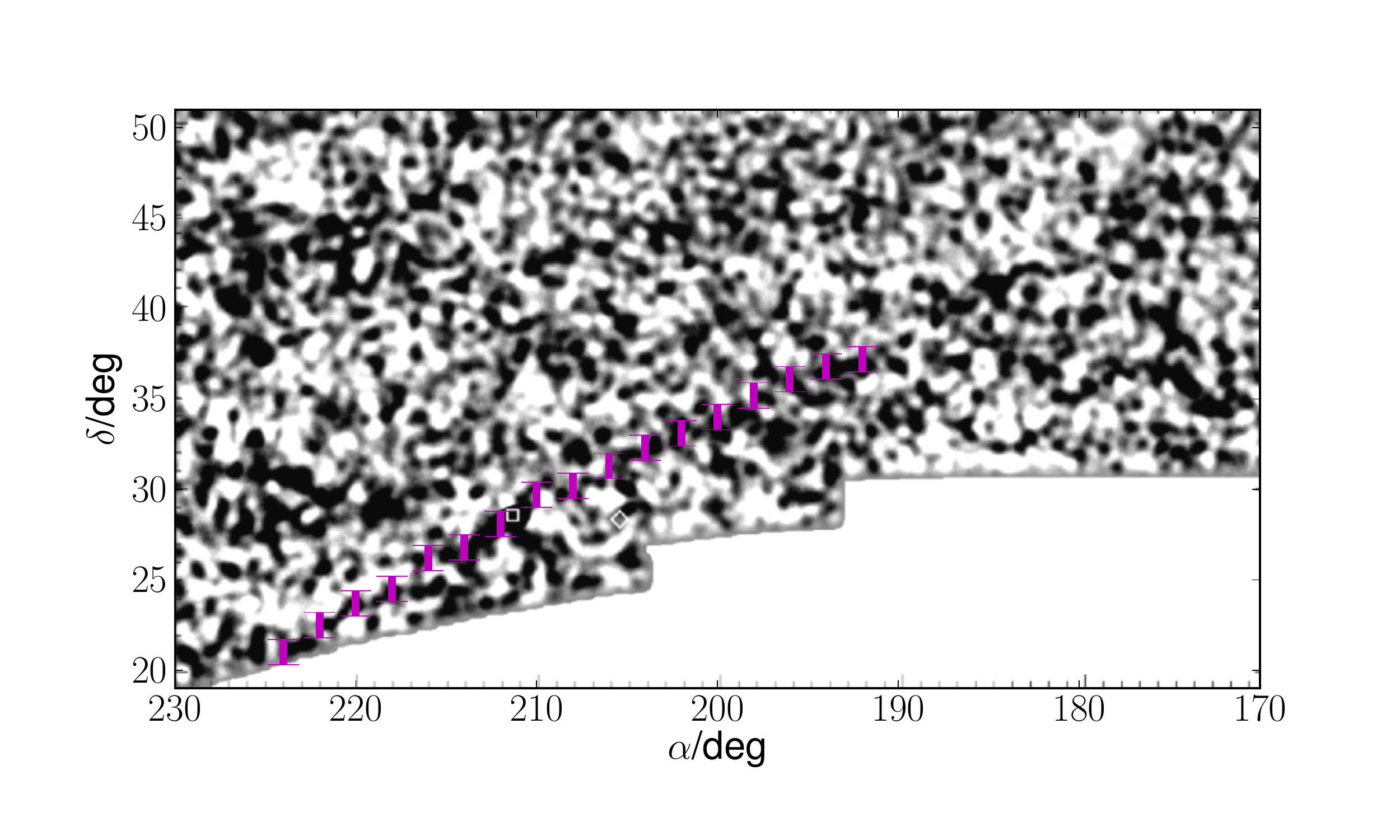}
\caption{The smoothed, summed, weighted image of the SDSS field for NGC 5466 after subtraction of a low order fit (background data taken from \citet{2006ApJ...639L..17G}, Figure 1). The magenta error bars mark the position of the stream that we consider as clearly connected with the globular cluster. The white square indicates the location of NGC 5466, while the white diamond shows the location of NGC 5272. We only trace the NGC 5466 stream for $\alpha\gtrsim192^\circ$, after which we consider its location to be ambiguous.}
\label{fig:Data}
\end{center}
\end{figure*}

\cite{2007MNRAS.380..749F} have modelled the stream numerically in a spherical halo potential including a disc and a bulge. They predict long ($\gtrsim 100^\circ$) and faint tidal tails for NGC 5466, confirming the extent and attenuation of the stream claimed by \cite{2006ApJ...639L..17G}. However, they are unable to reproduce the puzzling inconsistency with a smooth orbit as noted by \cite{2006ApJ...639L..17G} for the west end of the stream ($\alpha<190^\circ$). \cite{2006ApJ...639L..17G} speculate that this could owe to either irregularities in the Galactic potential, a recent encounter with a massive object or a confusion between streams. However, both the analysis by \cite{2007MNRAS.380..749F} and the orbits calculated by \cite{2006ApJ...639L..17G} assume a spherical shape for the Milky Way halo potential. In this work, we show that the deviation from a smooth orbit can be explained by a non-spherical/non-prolate halo potential.

This letter is organised as follows: In section \S\ref{sec:data} we describe the existing data and in \S\ref{sec:method} our method. The results are presented in \S\ref{sec:results} the results and discussed in \S\ref{sec:discuss}.

\section{Data}\label{sec:data}

For our analysis, we use the data provided by \citet{2006ApJ...639L..17G} as they report longer tails for NGC 5466 than \cite{2006ApJ...637L..29B}. Figure \ref{fig:Data} shows the data from their work with coloured error bars overlaid to indicate the positions of the stream as employed in this work\footnote{Note that the true extend of the stream is disputed. However, we assume in this work that the stream truly has the extend indicated in Figure \ref{fig:Data}.}. These positions range from $\alpha\in[192^\circ,224^\circ]$ as the stream leaves the area observed by SDSS at $\alpha=224^\circ$ and is considered ambiguous westward of $\alpha=192^\circ$. This ambiguous part of the stream corresponds to the deviation from a smooth orbit as noted by \cite{2006ApJ...639L..17G}  and is the subject of our investigation.

For the initial conditions of our orbit integration, we use the on-sky position (indicated by the white square in Fig. \ref{fig:Data}), distance and radial velocity for the globular cluster as given in the Harris catalogue \citep{1996AJ....112.1487H}, i.e. $\alpha$ = 14 05 27.29 hms, $\delta$ = 28 32 04.0 hms, $d =16.0\pm1.6\,$kpc, $v_r = 110.7$km/s. The Galactic centre distances is then 16.3 kpc. The proper motion data have been taken from \citep{2006ApJ...639L..17G}, i.e. [u,v,w] = [290, -240, 225]\,km/s. We assume $\pm 3 $mas/yr errors for the proper motions \citep{2004AJ....127.3034M}. Both the distances and the proper motions have relatively large errors in comparison to the angular position and radial velocities. They are varied within $3\sigma$ and $2\sigma$ errors, respectively. Note, that the orbital plane of NGC 5466 is roughly aligned with the Sun-Galactic centre axis and perpendicular to the disc, i.e. we observe the cluster from nearly within its orbital plane.

\section{Method}\label{sec:method}

\begin{figure*}
\centering
spherical\hspace{3.3cm} oblate \hspace{3.3cm} prolate \hspace{3.3cm} triaxial\\
\includegraphics[width=0.24\textwidth]{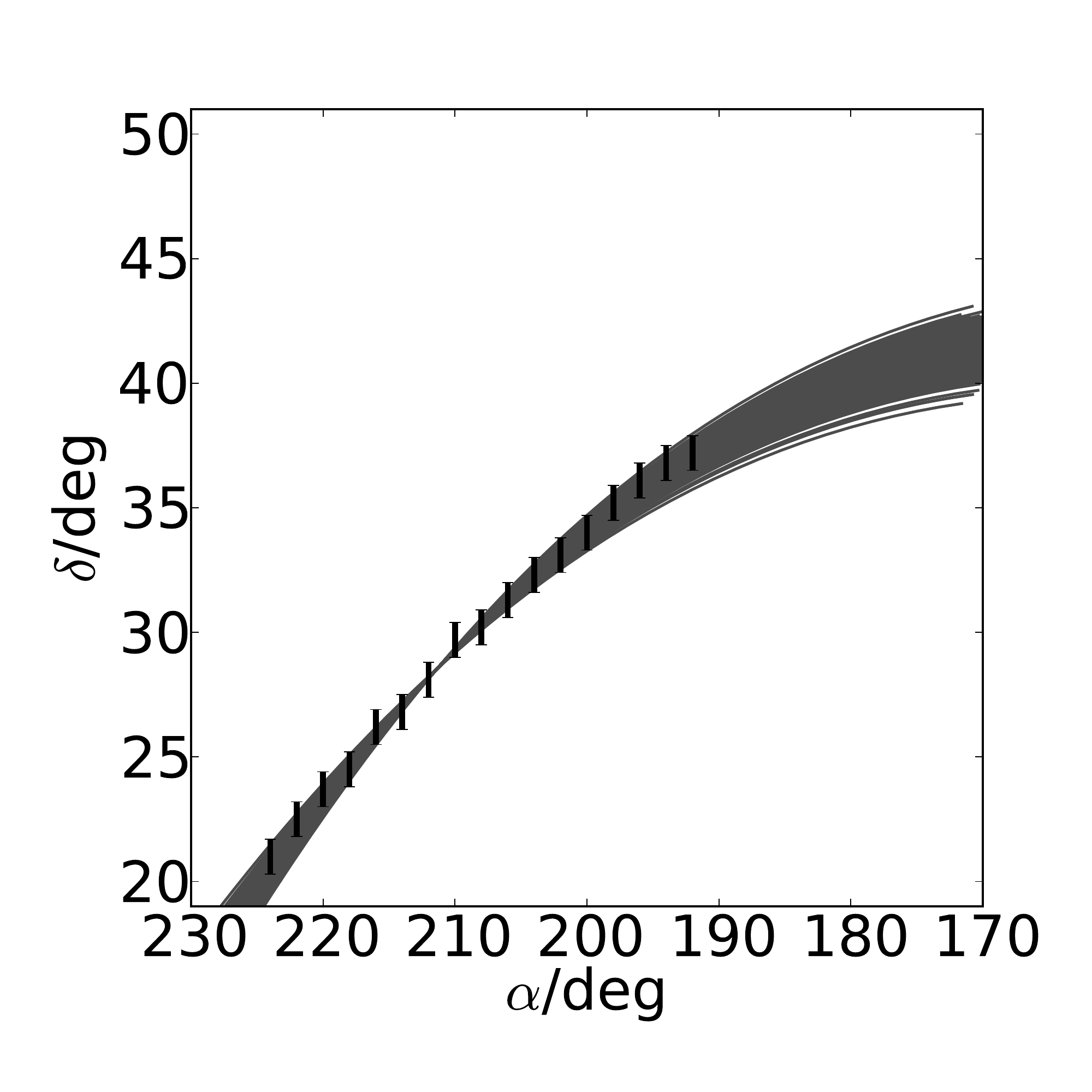}
\includegraphics[width=0.24\textwidth]{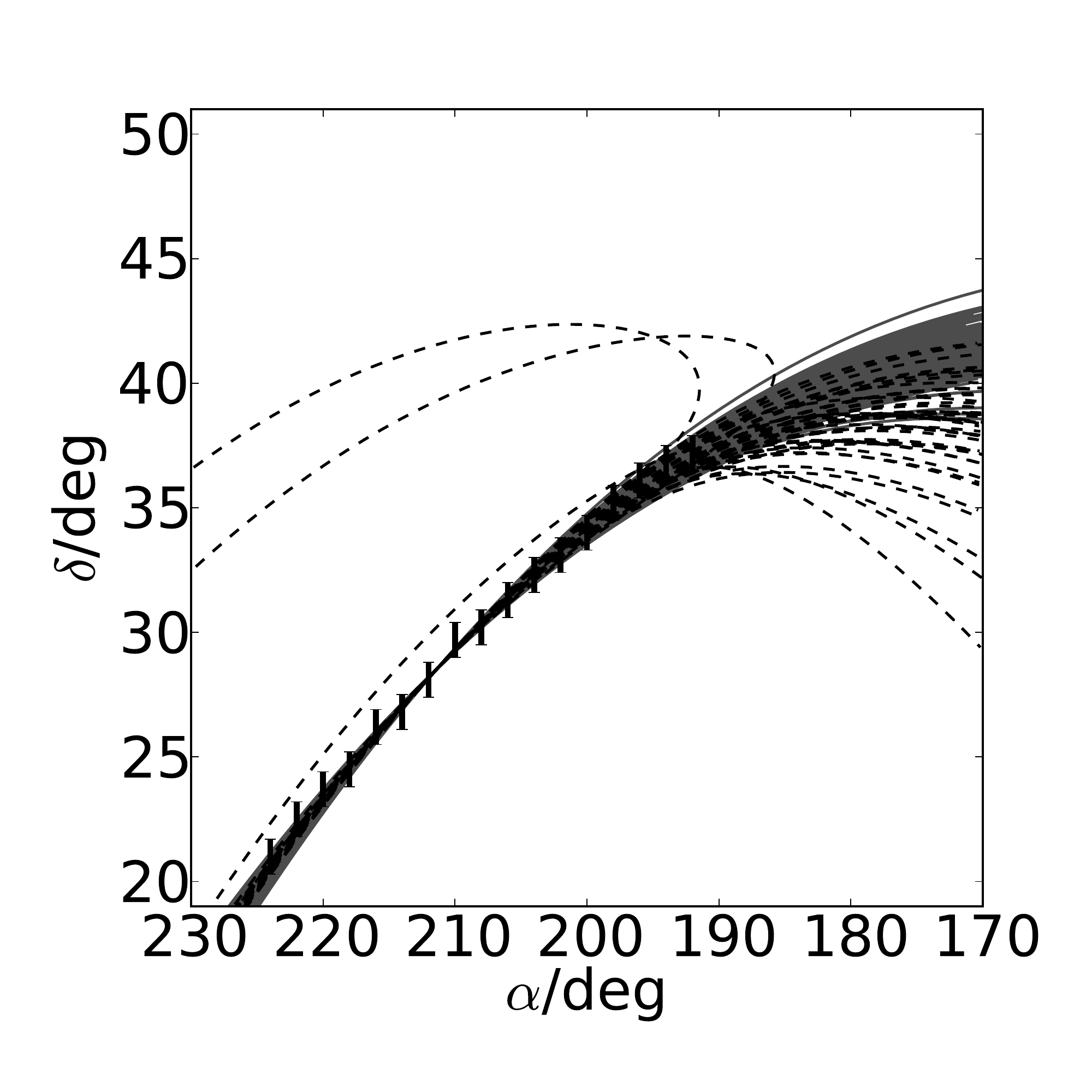}
\includegraphics[width=0.24\textwidth]{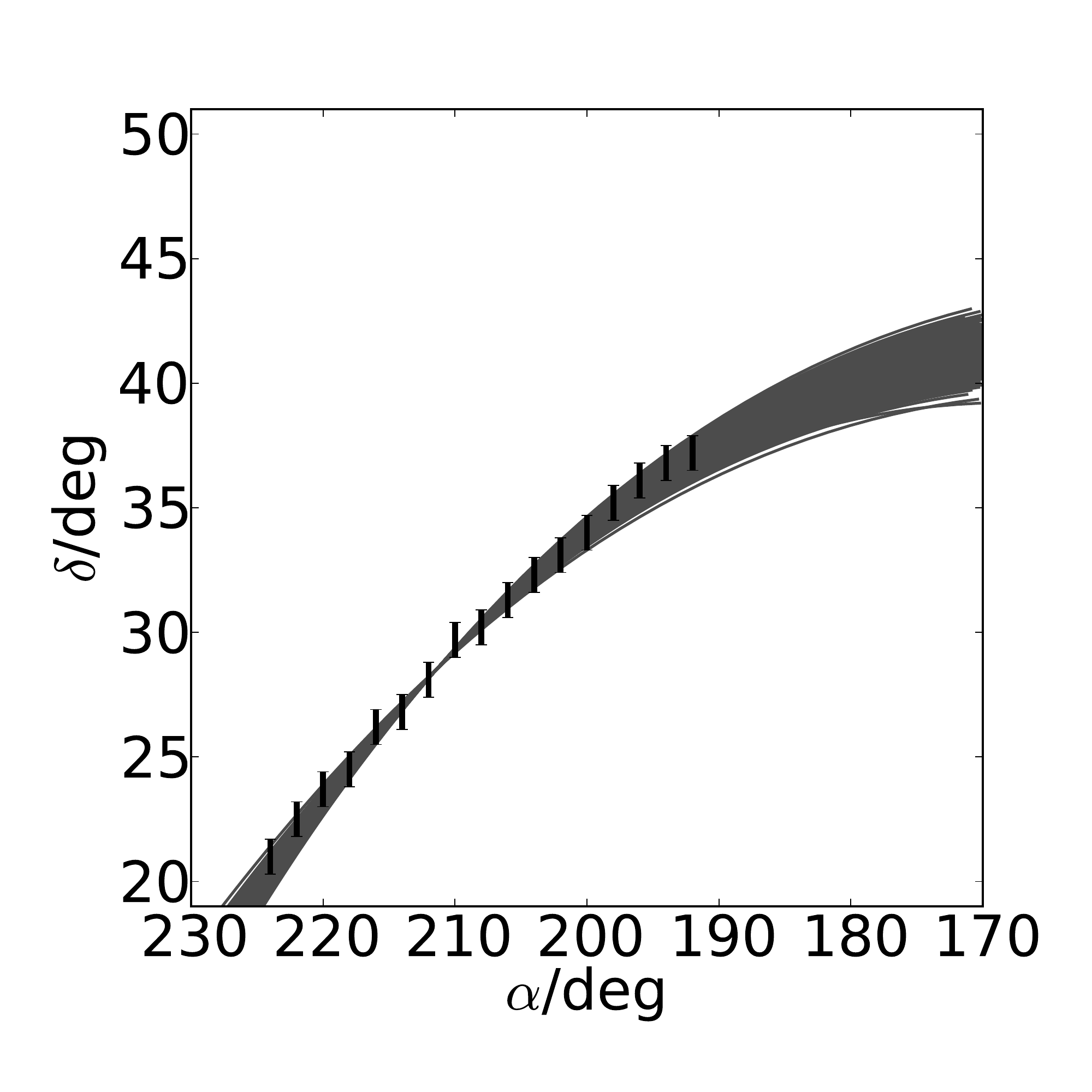}
\includegraphics[width=0.24\textwidth]{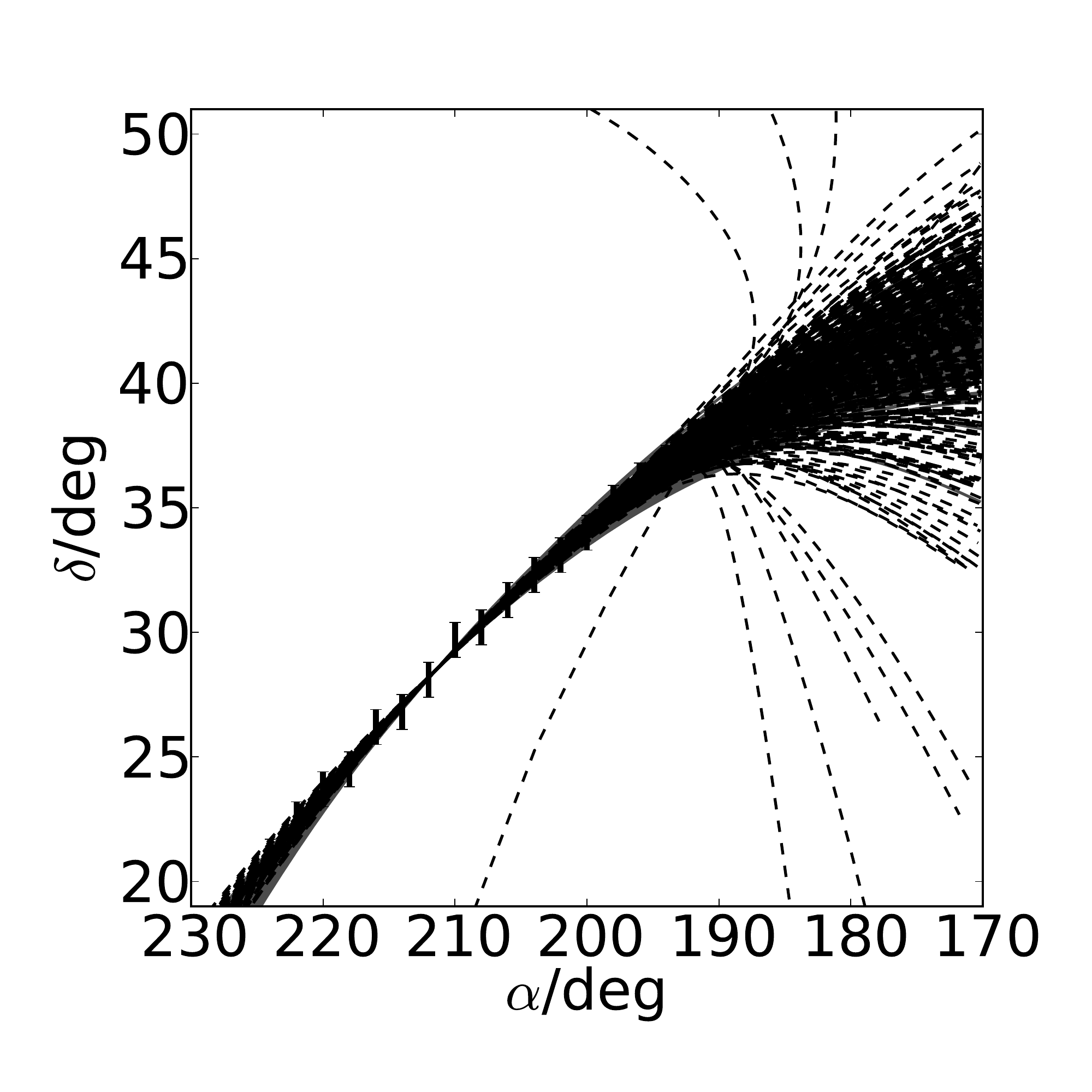}
\caption{The plots show the orbits of accepted models in the MCMC chain for the real NGC 5466 data and different halo shapes: Spherical ($q_1=q_z=1$), oblate ($q_1 = 1$, $\sqrt{1/2}\leq q_z \leq 0.95$), prolate ($q_1 = 1$, $1.05 \leq q_z \leq 1.8$), triaxial ($1.05 \leq q_1 \leq 1.8$, $\sqrt{1/2}\leq q_z \leq 1.8$). Orbits marked by black dashed lines have an apocentre in the Galactocentric coordinates, that appears within the field of view in projection. Note, that nearly all orbits with some deviation from a smooth orbit have an apocentre in projection within the field of view. Orbits without such a low apocentre are marked with grey lines. Note that we have omitted 90\% of these for clarity. The relative percentage of orbits with and without a low apocentre is not a converged quantity for the oblate and triaxial halo potentials and can vary widely depending on the individual MCMC chain (percentages $>50 \%$ are possible). However, we found no such `kink' orbits for the prolate halo potential, and just of order one out for 4 MCMC chains with $10^5$ samples for the spherical halo potential. (Note that the spherical halo potential model is still slightly oblate due to the presence of the disc.)}
\label{fig:NGC}
\end{figure*}

We analyse possible orbits of the NGC 5466 globular cluster that are consistent with the angular position of its tidal tails. Specifically, we investigate what effect the halo shape has on the angular positions of possible NGC 5466 orbits. 

The orbits of NGC 5466 are derived by integrating test particles from the position of the globular cluster forward and backwards in time. The initial conditions are taken from the globular cluster data as described in \S\ref{sec:data}, keeping the angular positions and radial velocities constant while varying the distance and proper motion data within their errors. We integrate the test particle orbits using the {\tt Orbit\_Int} code described in \cite{2010MNRAS.406.2312L}. 

We transform the Galactocentric coordinates into observable coordinates using M. Metz's tool {\tt bap.coords}\footnote{{\tt http://www.astro.uni-bonn.de/$\sim$mmetz/py/docs/mkj\_libs/ public/bap.coords-module.html}} \citep{2007MNRAS.374.1125M} adopting $8.0$\,kpc as the sun-Galactic centre distance and a circular velocity of 220\,km/s at the solar position.  

For our Milky Way potential model, we follow \cite{2010ApJ...714..229L} and use a \cite{1975PASJ...27..533M} disc
\begin{equation}
        \Phi_{\rm disc}=-  {GM_{\rm disc} \over
                 \sqrt{R^{2}+(a+\sqrt{z^{2}+b^{2}})^{2}}},
\end{equation}
a \cite{1990ApJ...356..359H} bulge
\begin{equation}
        \Phi_{\rm bulge}=-{GM_{\rm bulge} \over r+c},
\end{equation}
and a cored triaxial logarithmic potential
\begin{equation}
        \Phi_{\rm halo}=v_{\rm halo}^2 \ln (C_1 x^2 + C_2 y^2 +C_3 x y + (z/q_z)^2 + r_{\rm halo}^2)
\end{equation}
where the constants are defined as
\begin{equation}
C_1 = \left(\frac{\textrm{cos}^2 \phi}{q_{1}^2} + \frac{\textrm{sin}^2 \phi}{q_{2}^2}\right),
\end{equation}
\begin{equation}
C_2 = \left(\frac{\textrm{cos}^2 \phi}{q_{2}^2} + \frac{\textrm{sin}^2 \phi}{q_{1}^2}\right),
\end{equation}
\begin{equation}
C_3 = 2 \, \textrm{sin}\phi \, \textrm{cos} \phi \left( \frac{1}{q_{1}^2} - \frac{1}{q_{2}^2}\right).
\end{equation}
The disc mass is set to $M_{\rm disc}=1.0 \times 10^{11}$\,M$_{\odot}$ with a scale length $a=6.5$\,kpc and scale height $b=0.26$\,kpc. The mass of the bulge is fixed at $M_{\rm bulge}=3.4 \times 10^{10}$\,M$_{\odot}$ and its scale length $c=0.7$ kpc. The logarithmic potential is chosen in such a way, that the circular velocity of the total potential at the position of the sun ($R_\odot = 8$\,kpc) is equal to 220\,km/s. The core of the logarithmic potential is set by $r_{\rm halo} = 12$\,kpc and the shape parameter $q_2=1$ is kept fixed. By changing $q_1 \in [1,1.8]$, $\phi \in [0,\pi]$ and $q_z \in [1, 1.8]$ we can cover the whole range from purely spherical to fully triaxial halo potentials. Our results are not sensitive to this choice of potential functions and parameters (see discussion in section \ref{sec:results}).

We analyse orbits in spherical ($q_1=q_z=1$), oblate ($q_1 = 1$, $\sqrt{1/2}\leq q_z \leq 0.95$), prolate ($q_1 = 1$, $1.05 \leq q_z \leq 1.8$) and triaxial shaped halo potentials ($1.05 \leq q_1 \leq 1.8$, $\sqrt{1/2}\leq q_z \leq 1.8$). Note, that both the oblate and prolate halo potential have their symmetry axis aligned with the symmetry axis of the disc. All other configurations are included under the label ``triaxial''. In each potential we calculate orbits consistent with the tidal stream positions as marked in Figure \ref{fig:Data}. We use a Markov Chain Monte Carlo technique (MCMC)\footnote{The MCMC method is a probability distribution space sampling method employing a random walk to efficiently scan complex parameter spaces. The exact details of our implementation are described in Lux et al. (2012,  in prep).} with a varying step-size to effectively scan the parameter space varying the proper motions and distances of the initial conditions as well as the shape parameters, if appropriate.  

Note that only specific globular clusters streams with negligible stream-orbit-offsets can be analysed with the simple test particle fitting technique (Lux et al. 2012, in prep). Due to the special alignment of the NGC orbital plane with our position, we expect the stream offset from the orbit to be mainly in the distances and not in the angular positions. Then the expected stream-orbit-offset for this globular cluster stream is significantly smaller than the current distance measurement errors. However, here we do not aim to quantitatively constrain the individual shape parameters of the Milky Way halo, but simply point out the different types of orbits possible in various halo shapes. This should not be affected even if offsets in the angular positions occur. We test the convergence of our MCMC chains by running at least 4 chains with $10^5$ iterations each starting from different parameter sets.
 
\begin{figure}
\centering
\includegraphics[width=0.38\textwidth]{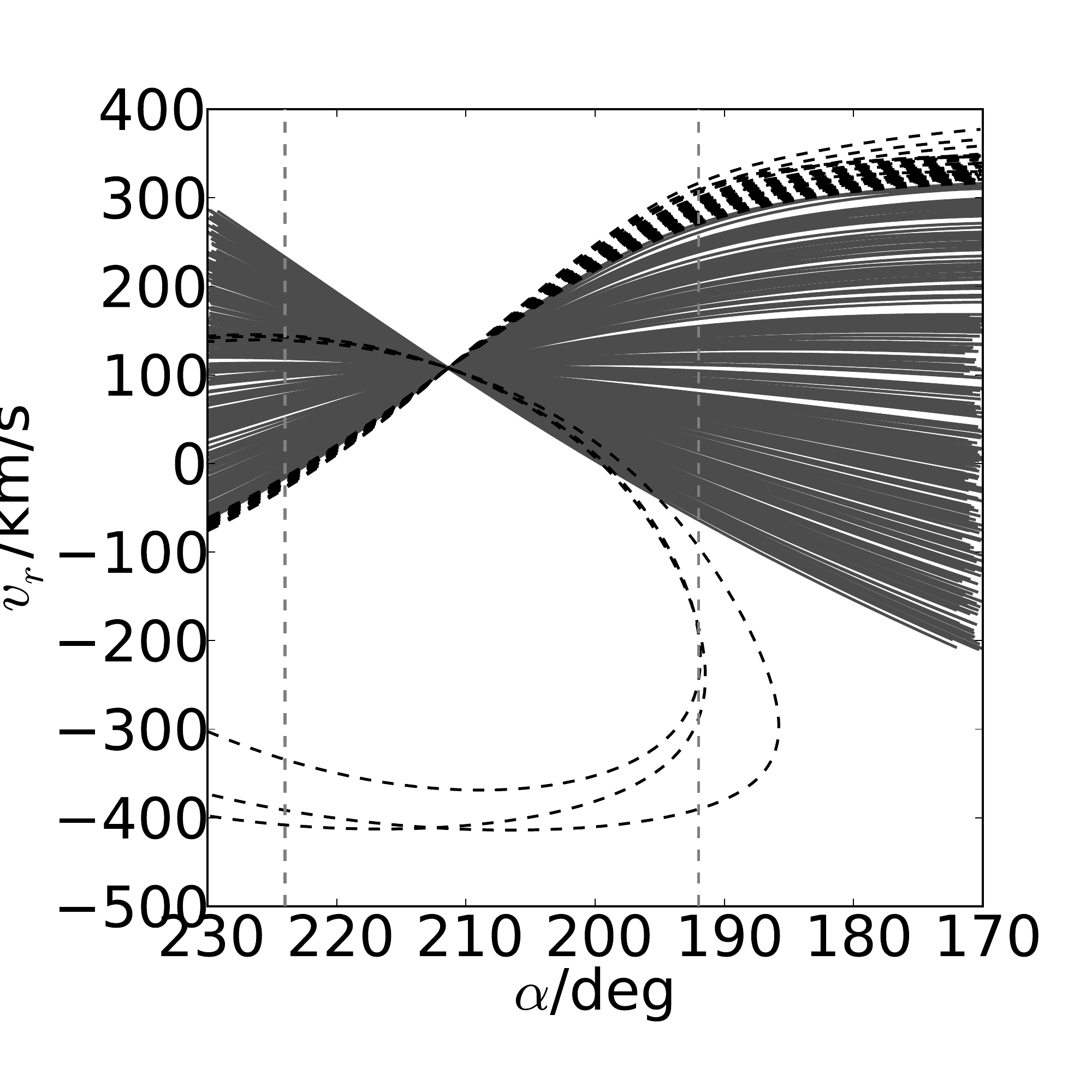}
\includegraphics[width=0.38\textwidth]{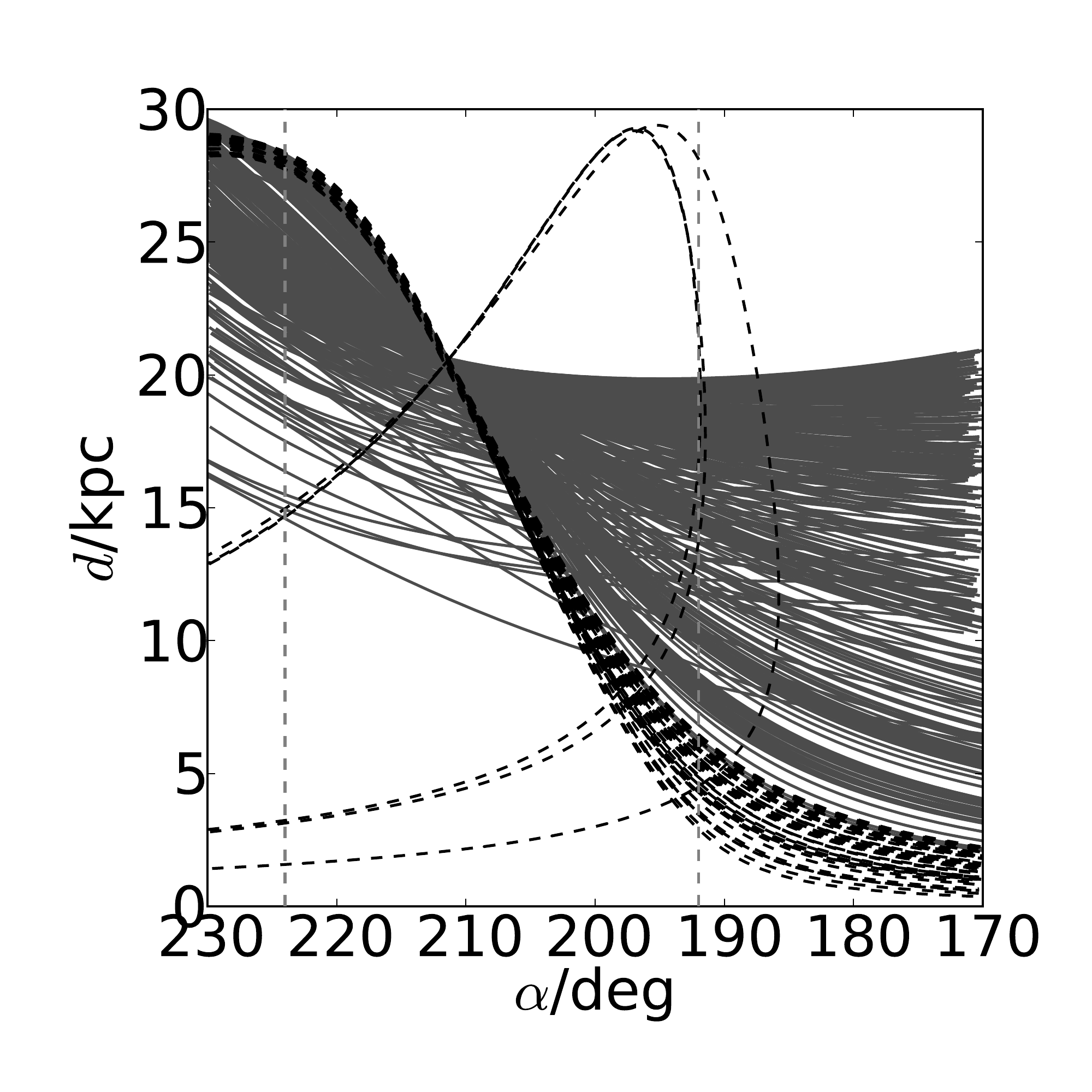}
\caption{The plots show the radial velocity data and heliocentric distances (assuming a Sun - Galactic centre distances of $8$kpc) of the accepted models in the MCMC chain for the oblate halo potential. The on-sky positions for these orbits are shown in Figure \ref{fig:NGC}. Orbits marked by black dashed lines have an apocentre in the Galactocentric coordinates that appears within the field of view in projection. Note that nearly all orbits with some deviation from a smooth orbit in the on-sky positions have an apocentre in projection within the field of view.  Orbits without such a low apocentre are marked with grey lines; we have omitted 90\% of these for clarity. The two dashed vertical lines mark the range of the NGC data as employed in this work.}
\label{fig:VRDIST}
\end{figure}

\section{Results} \label{sec:results}

Figure \ref{fig:NGC} shows the orbits consistent with the NGC 5466 stream data for the four different halo shapes: spherical ($q_1=q_z=1$), oblate ($q_1 = 1$, $\sqrt{1/2}\leq q_z \leq 0.95$), prolate ($q_1 = 1$, $1.05 \leq q_z \leq 1.8$) and triaxial ($1.05 \leq q_1 \leq 1.8$, $\sqrt{1/2}\leq q_z \leq 1.8$). We find that, while the spherical and prolate halo shapes only allow smooth orbits, in the oblate and triaxial halo shapes `kinky' orbits occur that can easily explain any deviations from a smooth orbit as seen by \citet{2006ApJ...639L..17G} at this end of the stream. Unfortunately, the deviations can either curve `upward' or `downward' making a prediction for the further location of the stream without any further data difficult. In both Figures \ref{fig:NGC} and \ref{fig:VRDIST} orbits with a low apocentre have been marked by black dashed lines. They show that these orbits not only distinguish themselves from the `smooth' orbits in angular positions but also in the distances and radial velocities along the stream.

To test the robustness of our results, we repeated our analysis using a triaxial NFW potential \citep[parametrisation as in][]{2009ApJ...702..890G} instead of the logarithmic potential for our halo (omitted for brevity). As for the logarithmic halo potential, only oblate and triaxial halo shapes allow the existence of ÕkinkyÕ orbits.

The existence of the disc potential already introduces a certain oblateness in the overall potential even for spherical halo shape. Therefore, one might suspect a degeneracy between disc and halo potential and expect to see `kinky' orbits even for fully spherical halos. However, we use a conservatively high disc mass \citep{2008gady.book.....B} and find the occurrence of a `kinky' orbit for a spherical halo shape negligible\footnote{Please note, that this might depend on the disputed extend of the stream and the specific data points chosen for this work.}. For the prolate halo shape, `kinky' orbits have never been found in our analysis. Therefore, the oblateness of the disc potential is not sufficient to mask the halo shape. Additionally, we tried replacing the Miyamoto-Nagai disc with an exponential disc to test whether this more realistic disc potential influences the oblateness of the total potential. We find that this is not the case and the amount of `kinky' orbits for spherical halos - given our assumed data -  is as negligible as for the Miyamoto-Nagai disc ($\sim 1$ occurrence in 4 MCMC chains with $10^5$ samples). 

Note, that the parameter space searched by the MCMC chains is degenerate due to the limited data set we are trying to fit. Therefore, the amount and the type of `kinky' orbits found for the oblate and triaxial halo potentials can vary significantly between individual chains. However, for the prolate halo potentials they are robustly consistent with zero, while for the spherical halo case such orbits are extremely rare. For the oblate and triaxial plots in Figure \ref{fig:NGC}, we decided to show MCMC chains with rather few `kinky' orbits as the less crowded plots are more instructive. Unfortunately, the nature of the `kink'  (even combined with distance or velocity information along the stream)  does not allow a distinction between the oblate and triaxial halo potentials. 

We find that the low apocentres are 8-15 kpc away from the globular cluster (3D distance). \cite{2007ApJ...659.1212M} find that GCs on eccentric orbits near apocentre are expected to have multiple tails around the globular cluster. This could create further deviations from a smooth stream. Additionally, streams at apocentre are compressed in both angular positions and radial velocities and therefore might yield poorer constraints on the underlying potential than stream further away from apocentre \citep{2009MNRAS.400..548E}.   However, a lower apocentre also means a lower eccentricity and \cite{2008ApJ...689..936J} find a more stream-like morphology for less eccentric orbits. As low apocentres only exist in oblate/triaxial halo potentials, this does not effect our results. Testing the effects of a close apocentre on the stream morphology with N-body models is therefore beyond the scope of this paper and will be postponed to future work.

\section{Discussion and Conclusions} \label{sec:discuss}

We investigate a possible explanation of the apparent deviation from a smooth orbit of the globular cluster tidal stream NGC 5466 westward of $\alpha\gtrsim192^\circ$. We integrate orbits in a variety of halo shapes consistent with the known location of the stream and find that only for either oblate halo potentials (with respect to the disc) or fully triaxial halo potentials such deviations are possible. If this deviation is verified, it places a strong constraint on the shape of our Milky Way halo and both spherical and prolate halo shapes can be excluded with high probability.

The observed deviations are highly correlated with low apocentres in the globular cluster orbits and can curve either way from the current orbit. This makes a prediction for the halo shape without additional data difficult. However, these low apocentre orbits not only show a very distinct behaviour in on-sky positions, but also in radial velocities/distances along the stream. Therefore, further maps of the westward end of the stream as well as radial velocities/distances along the whole known stream are the key to unraveling the Milky Way halo shape using the NGC 5466 stream.

\section*{Acknowledgments}

The authors acknowledge the valuable comments of the anonymous referee. HL gratefully acknowledges helpful discussions with Frazer Pearce, Steven Bamford and Mike Merrifield. Furthermore, HL acknowledges a fellowship from the European CommissionÕs Framework Programme 7, through the Marie Curie Initial Training Network CosmoComp (PITN-GA-2009-238356). JIR would like to acknowledge support from SNF grant PP00P2\_128540 / 1. 

\bibliographystyle{mn2e}
\bibliography{NGCLetter}

\begin{thebibliography}{}

\bibitem[\protect\citeauthoryear{{Belokurov}, {Evans}, {Irwin}, {Hewett} \&
  {Wilkinson}}{{Belokurov} et~al.}{2006}]{2006ApJ...637L..29B}
{Belokurov} V.,  {Evans} N.~W.,  {Irwin} M.~J.,  {Hewett} P.~C.,    {Wilkinson}
  M.~I.,  2006, \apjl, 637, L29

\bibitem[\protect\citeauthoryear{{Belokurov}, {Zucker}, {Evans}, {Gilmore},
  {Vidrih}, {Bramich}, {Newberg}, {Wyse} \& {et al.}}{{Belokurov}
  et~al.}{2006}]{2006ApJ...642L.137B}
{Belokurov} V.,  {Zucker} D.~B.,  {Evans} N.~W.,  {Gilmore} G.,  {Vidrih} S.,
  {Bramich} D.~M.,  {Newberg} H.~J.,  {Wyse} R.~F.~G.,    {et al.} 2006, \apjl,
  642, L137

\bibitem[\protect\citeauthoryear{{Binney} \& {Tremaine}}{{Binney} \&
  {Tremaine}}{2008}]{2008gady.book.....B}
{Binney} J.,  {Tremaine} S.,  2008, {Galactic Dynamics: Second Edition}.
Galactic Dynamics: Second Edition, by James Binney and Scott Tremaine.~ISBN
  978-0-691-13026-2 (HB).~Published by Princeton University Press, Princeton,
  NJ USA, 2008.

\bibitem[\protect\citeauthoryear{{Debattista}, {Moore}, {Quinn}, {Kazantzidis},
  {Maas}, {Mayer}, {Read} \& {Stadel}}{{Debattista}
  et~al.}{2008}]{2008ApJ...681.1076D}
{Debattista} V.~P.,  {Moore} B.,  {Quinn} T.,  {Kazantzidis} S.,  {Maas} R.,
  {Mayer} L.,  {Read} J.,    {Stadel} J.,  2008, \apj, 681, 1076

\bibitem[\protect\citeauthoryear{{Dubinski}}{{Dubinski}}{1994}]{1994ApJ...431.%
.617D}
{Dubinski} J.,  1994, \apj, 431, 617

\bibitem[\protect\citeauthoryear{{Dubinski} \& {Carlberg}}{{Dubinski} \&
  {Carlberg}}{1991}]{1991ApJ...378..496D}
{Dubinski} J.,  {Carlberg} R.~G.,  1991, \apj, 378, 496

\bibitem[\protect\citeauthoryear{{Eyre} \& {Binney}}{{Eyre} \&
  {Binney}}{2009}]{2009MNRAS.400..548E}
{Eyre} A.,  {Binney} J.,  2009, \mnras, 400, 548

\bibitem[\protect\citeauthoryear{{Fellhauer}, {Belokurov}, {Evans},
  {Wilkinson}, {Zucker}, {Gilmore}, {Irwin}, {Bramich}, {Vidrih}, {Wyse},
  {Beers} \& {Brinkmann}}{{Fellhauer} et~al.}{2006}]{2006ApJ...651..167F}
{Fellhauer} M.,  {Belokurov} V.,  {Evans} N.~W.,  {Wilkinson} M.~I.,  {Zucker}
  D.~B.,  {Gilmore} G.,  {Irwin} M.~J.,  {Bramich} D.~M.,  {Vidrih} S.,  {Wyse}
  R.~F.~G.,  {Beers} T.~C.,    {Brinkmann} J.,  2006, \apj, 651, 167

\bibitem[\protect\citeauthoryear{{Fellhauer}, {Evans}, {Belokurov}, {Wilkinson}
  \& {Gilmore}}{{Fellhauer} et~al.}{2007}]{2007MNRAS.380..749F}
{Fellhauer} M.,  {Evans} N.~W.,  {Belokurov} V.,  {Wilkinson} M.~I.,
  {Gilmore} G.,  2007, \mnras, 380, 749

\bibitem[\protect\citeauthoryear{{Grillmair} \& {Dionatos}}{{Grillmair} \&
  {Dionatos}}{2006}]{2006ApJ...641L..37G}
{Grillmair} C.~J.,  {Dionatos} O.,  2006, \apjl, 641, L37

\bibitem[\protect\citeauthoryear{{Grillmair} \& {Johnson}}{{Grillmair} \&
  {Johnson}}{2006}]{2006ApJ...639L..17G}
{Grillmair} C.~J.,  {Johnson} R.,  2006, \apjl, 639, L17

\bibitem[\protect\citeauthoryear{{Guedes}, {Madau}, {Kuhlen}, {Diemand} \&
  {Zemp}}{{Guedes} et~al.}{2009}]{2009ApJ...702..890G}
{Guedes} J.,  {Madau} P.,  {Kuhlen} M.,  {Diemand} J.,    {Zemp} M.,  2009,
  \apj, 702, 890

\bibitem[\protect\citeauthoryear{{Harris}}{{Harris}}{1996}]{1996AJ....112.1487%
H}
{Harris} W.~E.,  1996, \aj, 112, 1487

\bibitem[\protect\citeauthoryear{{Helmi}}{{Helmi}}{2004}]{2004ApJ...610L..97H}
{Helmi} A.,  2004, \apjl, 610, L97

\bibitem[\protect\citeauthoryear{{Hernquist}}{{Hernquist}}{1990}]{1990ApJ...35%
6..359H}
{Hernquist} L.,  1990, \apj, 356, 359

\bibitem[\protect\citeauthoryear{{Ibata}, {Irwin}, {Lewis} \& {Stolte}}{{Ibata}
  et~al.}{2001}]{2001ApJ...547L.133I}
{Ibata} R.,  {Irwin} M.,  {Lewis} G.~F.,    {Stolte} A.,  2001, \apjl, 547,
  L133

\bibitem[\protect\citeauthoryear{{Ibata}, {Lewis}, {Irwin}, {Totten} \&
  {Quinn}}{{Ibata} et~al.}{2001}]{2001ApJ...551..294I}
{Ibata} R.,  {Lewis} G.~F.,  {Irwin} M.,  {Totten} E.,    {Quinn} T.,  2001,
  \apj, 551, 294

\bibitem[\protect\citeauthoryear{{Jing} \& {Suto}}{{Jing} \&
  {Suto}}{2002}]{2002ApJ...574..538J}
{Jing} Y.~P.,  {Suto} Y.,  2002, \apj, 574, 538

\bibitem[\protect\citeauthoryear{{Johnston}, {Bullock}, {Sharma}, {Font},
  {Robertson} \& {Leitner}}{{Johnston} et~al.}{2008}]{2008ApJ...689..936J}
{Johnston} K.~V.,  {Bullock} J.~S.,  {Sharma} S.,  {Font} A.,  {Robertson}
  B.~E.,    {Leitner} S.~N.,  2008, \apj, 689, 936

\bibitem[\protect\citeauthoryear{{Johnston}, {Law} \& {Majewski}}{{Johnston}
  et~al.}{2005}]{2005ApJ...619..800J}
{Johnston} K.~V.,  {Law} D.~R.,    {Majewski} S.~R.,  2005, \apj, 619, 800

\bibitem[\protect\citeauthoryear{{Kazantzidis}, {Abadi} \&
  {Navarro}}{{Kazantzidis} et~al.}{2010}]{2010ApJ...720L..62K}
{Kazantzidis} S.,  {Abadi} M.~G.,    {Navarro} J.~F.,  2010, \apjl, 720, L62

\bibitem[\protect\citeauthoryear{{Koposov}, {Rix} \& {Hogg}}{{Koposov}
  et~al.}{2010}]{2010ApJ...712..260K}
{Koposov} S.~E.,  {Rix} H.,    {Hogg} D.~W.,  2010, \apj, 712, 260

\bibitem[\protect\citeauthoryear{{Law}, {Johnston} \& {Majewski}}{{Law}
  et~al.}{2005}]{2005ApJ...619..807L}
{Law} D.~R.,  {Johnston} K.~V.,    {Majewski} S.~R.,  2005, \apj, 619, 807

\bibitem[\protect\citeauthoryear{{Law} \& {Majewski}}{{Law} \&
  {Majewski}}{2010}]{2010ApJ...714..229L}
{Law} D.~R.,  {Majewski} S.~R.,  2010, \apj, 714, 229

\bibitem[\protect\citeauthoryear{{Law}, {Majewski} \& {Johnston}}{{Law}
  et~al.}{2009}]{2009ApJ...703L..67L}
{Law} D.~R.,  {Majewski} S.~R.,    {Johnston} K.~V.,  2009, \apjl, 703, L67

\bibitem[\protect\citeauthoryear{{Lin} \& {Lynden-Bell}}{{Lin} \&
  {Lynden-Bell}}{1977}]{1977MNRAS.181...59L}
{Lin} D.~N.~C.,  {Lynden-Bell} D.,  1977, \mnras, 181, 59

\bibitem[\protect\citeauthoryear{{Lux}, {Read} \& {Lake}}{{Lux}
  et~al.}{2010}]{2010MNRAS.406.2312L}
{Lux} H.,  {Read} J.~I.,    {Lake} G.,  2010, \mnras, 406, 2312

\bibitem[\protect\citeauthoryear{{Metz}, {Kroupa} \& {Jerjen}}{{Metz}
  et~al.}{2007}]{2007MNRAS.374.1125M}
{Metz} M.,  {Kroupa} P.,    {Jerjen} H.,  2007, \mnras, 374, 1125

\bibitem[\protect\citeauthoryear{{Miyamoto} \& {Nagai}}{{Miyamoto} \&
  {Nagai}}{1975}]{1975PASJ...27..533M}
{Miyamoto} M.,  {Nagai} R.,  1975, \pasj, 27, 533

\bibitem[\protect\citeauthoryear{{Montuori}, {Capuzzo-Dolcetta}, {Di Matteo},
  {Lepinette} \& {Miocchi}}{{Montuori} et~al.}{2007}]{2007ApJ...659.1212M}
{Montuori} M.,  {Capuzzo-Dolcetta} R.,  {Di Matteo} P.,  {Lepinette} A.,
  {Miocchi} P.,  2007, \apj, 659, 1212

\bibitem[\protect\citeauthoryear{{Munn}, {Monet}, {Levine}, {Canzian}, {Pier},
  {Harris}, {Lupton}, {Ivezi{\'c}}, {Hindsley}, {Hennessy}, {Schneider} \&
  {Brinkmann}}{{Munn} et~al.}{2004}]{2004AJ....127.3034M}
{Munn} J.~A.,  {Monet} D.~G.,  {Levine} S.~E.,  {Canzian} B.,  {Pier} J.~R.,
  {Harris} H.~C.,  {Lupton} R.~H.,  {Ivezi{\'c}} {\v Z}.,  {Hindsley} R.~B.,
  {Hennessy} G.~S.,  {Schneider} D.~P.,    {Brinkmann} J.,  2004, \aj, 127,
  3034

\bibitem[\protect\citeauthoryear{{Newberg}, {Willett}, {Yanny} \&
  {Xu}}{{Newberg} et~al.}{2010}]{2010ApJ...711...32N}
{Newberg} H.~J.,  {Willett} B.~A.,  {Yanny} B.,    {Xu} Y.,  2010, \apj, 711,
  32

\bibitem[\protect\citeauthoryear{{Odenkirchen}, {Grebel}, {Dehnen}, {Rix},
  {Yanny}, {Newberg}, {Rockosi}, {Mart{\'{\i}}nez-Delgado}, {Brinkmann} \&
  {Pier}}{{Odenkirchen} et~al.}{2003}]{2003AJ....126.2385O}
{Odenkirchen} M.,  {Grebel} E.~K.,  {Dehnen} W.,  {Rix} H.,  {Yanny} B.,
  {Newberg} H.~J.,  {Rockosi} C.~M.,  {Mart{\'{\i}}nez-Delgado} D.,
  {Brinkmann} J.,    {Pier} J.~R.,  2003, \aj, 126, 2385

\bibitem[\protect\citeauthoryear{{Pe{\~n}arrubia}, {Belokurov}, {Evans},
  {Mart{\'{\i}}nez-Delgado}, {Gilmore}, {Irwin}, {Niederste-Ostholt} \&
  {Zucker}}{{Pe{\~n}arrubia} et~al.}{2010}]{2010MNRAS.408L..26P}
{Pe{\~n}arrubia} J.,  {Belokurov} V.,  {Evans} N.~W.,
  {Mart{\'{\i}}nez-Delgado} D.,  {Gilmore} G.,  {Irwin} M.,
  {Niederste-Ostholt} M.,    {Zucker} D.~B.,  2010, \mnras, 408, L26

\bibitem[\protect\citeauthoryear{{Pe{\~n}arrubia}, {Zucker}, {Irwin}, {Hyde},
  {Lane}, {Lewis}, {Gilmore}, {Wyn Evans} \& {Belokurov}}{{Pe{\~n}arrubia}
  et~al.}{2011}]{2011ApJ...727L...2P}
{Pe{\~n}arrubia} J.,  {Zucker} D.~B.,  {Irwin} M.~J.,  {Hyde} E.~A.,  {Lane}
  R.~R.,  {Lewis} G.~F.,  {Gilmore} G.,  {Wyn Evans} N.,    {Belokurov} V.,
  2011, \apjl, 727, L2+

\bibitem[\protect\citeauthoryear{{Read} \& {Moore}}{{Read} \&
  {Moore}}{2005}]{2005MNRAS.361..971R}
{Read} J.~I.,  {Moore} B.,  2005, \mnras, 361, 971

\bibitem[\protect\citeauthoryear{{Varghese}, {Ibata} \& {Lewis}}{{Varghese}
  et~al.}{2011}]{2011MNRAS.417..198V}
{Varghese} A.,  {Ibata} R.,    {Lewis} G.~F.,  2011, \mnras, 417, 198

\bibitem[\protect\citeauthoryear{{Willett}, {Newberg}, {Zhang}, {Yanny} \&
  {Beers}}{{Willett} et~al.}{2009}]{2009ApJ...697..207W}
{Willett} B.~A.,  {Newberg} H.~J.,  {Zhang} H.,  {Yanny} B.,    {Beers} T.~C.,
  2009, \apj, 697, 207

\end{thebibliography}

\end{document}